\documentclass[aps,prc,nofootinbib]{revtex4}
\usepackage{eurosym}
\usepackage{amsmath,amssymb}
\usepackage{graphicx}
\usepackage{dcolumn}
\usepackage{bm}
\usepackage{color}
\usepackage{multirow}
\usepackage{float}
\usepackage{array}
\usepackage{dcolumn}
\usepackage{booktabs}
\usepackage[detect-all]{siunitx}
\usepackage{makecell}
\usepackage{xcolor}
\usepackage{yfonts}

\graphicspath{ {/Users/Nastia/Desktop/research_dis/papers_writing/xenes/nov_5} }

\newcolumntype{P}[1]{>{\centering\arraybackslash}p{#1}}
\begin{document}

\title{Folding procedure for $\Omega$-$\alpha$ potential}
\author{I. Filikhin$^1$, R. Ya. Kezerashvili$^{2,3,4}$, and B. Vlahovic$^1$}
\affiliation{\mbox{$^{1}$North Carolina Central University, Durham, NC, USA} \\
$^{2}$New York City College of Technology, The City University of New York,
Brooklyn, NY, USA\\
$^{3}$The Graduate School and University Center, The City University of New
York, New York, NY, USA\\
$^{4}$Long Island University, Brooklyn, NY, USA}

\begin{abstract}
\noindent Using the folding procedure, we investigate the bound state of the $\Omega$+$\alpha$ system based on $\Omega$-$N$ ($^{5}S_{2}$) HAL QCD potential. Previous theoretical analyses have indicated the existence of a deeply bound ground state, which is attributed to the strong $\Omega$-nucleon interaction. By employing well-established parameterizations of nucleon density within the alpha particle, and the central HAL QCD $\Omega$-$N$ potential, we performed numerical calculations for the folding $\Omega$-$\alpha$ potential. Our results show that the $V_{\Omega\alpha}(r)$ potential can be accurately fitted using a Woods-Saxon function, with a phenomenological parameter $R = 1.1A^{1/3} \approx 1.74$ fm ($A=4$) in the asymptotic region where $2 < r < 3$ fm. We provide a thorough description of the corresponding numerical procedure. Our evaluation of the binding energy of the $\Omega$+$\alpha$ system within the cluster model is consistent with both previous and recent reported findings. To further validate the folding procedure, we also calculated the $\Xi$-$\alpha$ folding potential based on a simulation of the ESC08c $Y$-$N$ Nijmegen model. A comprehensive comparison between the $\Xi$-$\alpha$ folding and $\Xi$-$ \alpha$ phenomenological potentials is presented and discussed.

\end{abstract}

\maketitle
\date{\today }

\section{Introduction}

The omega baryons, $\Omega $, are a family of hadron particles that are
either neutral or have a +2, +1 or -1 elementary charge. Negatively charged
omega, $\Omega ^{-}(sss)$, are made of three strange quarks \cite%
{B64,DataGroup,2009} and has a rest mass 1672.45 MeV/c$^{2}$ \cite%
{Partlist2012}. 
The $\Omega N$ dibaryon was predicted to be bound in various quark model calculations \cite{Goldman1987,Oka1988,Li2000,Pang2004}. The existence of a bound $\Omega N$ dibaryon state with strangeness $-3$ was first suggested in Ref. \cite{Goldman1987} using potential quark and MIT bag models. Possible candidates for $S$-wave dibaryons with different strange numbers, including $\Omega N$, were studied within the chiral $SU(3)$ quark model \cite{Li2000}. The binding energy of the six-quark system with strangeness $s = -3$ was investigated under the chiral $SU(3)$ constituent quark model using the Resonating Group Method, focusing on the single $\Omega N$ channel with spin $S = 2$. The effective $\Omega$-$N$ interaction was explored in the refined Quark Delocalization Color Screening Model (QDCSM) \cite{Pang2004}. The bound states are possible due to their unique structure, which has minimal contributions from the color-magnetic interaction. Further studies of the $\Omega N$ dibaryon within the QDCSM and the chiral quark model were conducted in Refs. \cite{Zhu2015,Huang2015}. A lattice QCD analysis with nearly physical quark masses was performed in Refs. \cite{Etminan2014,Morita2016,LagrangianMethod,Iritani2019}. A formalism for treating the scattering of decuplet baryons in chiral effective field theory was developed, providing the minimal Lagrangian and potentials in leading-order $SU(3)$ chiral effective field theory for the interactions of octet and decuplet baryons \cite{Meissner2017}. This formalism was applied to $\Omega N$ and $\Omega \Omega$ scattering, and the results were compared with lattice QCD simulations. Although the details of these studies differ, they consistently indicate the existence of the $\Omega N$ bound state.
Using the $\Omega$-$N$ local potential from Ref. \cite{LagrangianMethod}, the authors studied the $\Omega d$ system in the maximal spin channel $(I)J^{P}=(0)5/2^{+}$ \cite{GV0}, while in Ref. \cite{GV} calculations were performed for the $\Omega NN$ system using the HAL QCD Collaboration interaction \cite{Iritani2019}. It is worth noting that recently, the first results on the interaction between the $\phi$-meson and the nucleon have been presented based on
(2+1)-flavor lattice QCD simulations with nearly physical quark masses \cite{Lyu22}. This interaction was used to study the $\phi NN$ system \cite{EA24,FKVPRD2024} and was employed in \cite{FilKez2024} to calculate binding energies of the hypothetical $\phi$+$\alpha$ and $\phi$+$\alpha$+$\alpha$ systems.

The nuclear $\Omega N$ system in the $S$-wave and spin-2 channel (${^5}S_2$) is studied from the (2+1)-flavor lattice QCD with nearly physical quark masses in Ref. \cite{Iritani2019}. The $\Omega$-$N$ potential proposed in Ref. \cite{Iritani2019}, was employed in \cite{Etminan2020} to calculate the $\Omega$+$\alpha$+$\alpha$ system binding energy. To achieve this, the $\Omega$-$\alpha$ potential was determined using a folding procedure. The $\Omega \alpha$ presents a five-particle system. A reduction of a five-body problem to an effective two-body system has computational advantages. One can temp 
to make such a reduction for $\Omega \alpha$ system, which, approximately, can be considered as $\Omega$-particle and $^4$He nucleus that 
appears to be very well suited for a two-body description, as previously attempted. The first step of the procedure is obviously to construct the $\Omega$-$\alpha$
effective potential by folding the $\alpha$-particle nuclear density distribution
with the $\Omega$-$N$ interaction. The remaining steps are computationally much
simpler due to the two-body nature calculations. 

It is essential to note that folding procedures typically require significant adjustments to the interactions incorporated into the method \cite{double}. The procedure proposed in Ref. \cite{Etminan2020} addresses this issue by employing the asymptotic region to carry out these adjustments and a Woods-Saxon (WS) function was utilized to fit the $\Omega$-$\alpha$ potential in the asymptotic region, starting at approximately 2 fm.

In the present work, we have carefully replicated the folding procedure proposed in Refs. \cite{Etminan2020} to ensure that the obtained $\Omega$-$\alpha$ potential reliably describes the interaction in the asymptotic region. Our focus is on the numerical aspects of the folding process, where we highlighted the uncertainties that arise from the assumptions underlying this asymptotic approach used in the method. 
This article has two foci: i. Construction of a $\Omega$-$\alpha$ interaction based on the HAL QCD $\Omega$-$N$ interaction \cite{Iritani2019}; ii. study the possible formation of $^{5}_{\Omega}$He hypernuclei as the $\Omega$+$\alpha$ system in the framework
of a two-body cluster model.

This paper is organized as follows: First, in Sec. \ref{Folding} we present the folding procedure of the HAL QCD $\Omega$-$N$ interaction in
the ${^5}S_2$ channel with the matter distributions of $^4$He that reproduce the experimental root-mean-square radius of this nucleus and propose the $\Omega$-$\alpha$ effective potential obtained based on the Wood-Saxson fit of the folding of the HAL QCD $\Omega$-$N$ interaction. 
In Sec. \ref{Calculations} we perform numerical calculations for the folding $\Omega$-$\alpha$ potential, provide its fit using a WS function and present the binding energy (BE) of the $^{5}_{\Omega}$He hypernucleus.  
Section \ref{Summary} is devoted to the concluding remarks of this study.

\section{$\Omega$-$\alpha$ folding potential}

\label{Folding}

Investigations of the $\Omega\alpha$ system properties within a nonrelativistic potential model require a knowledge of the corresponding $\Omega$-$N$
 interaction potential.
Recently, in Ref. \cite{Iritani2019} $\Omega N$ dibaryon in
the $S$-wave and spin-2 channel is studied from the (2+1)-flavor lattice QCD
with nearly physical quark masses ($m_{\pi }=$ 146 MeV and $m_{K}=$ 525MeV)
by employing the HAL QCD method. The $\Omega$-$ N$ ($^{5}S_{2}$) potential,
obtained under the assumption that its couplings to the $D-$wave
octet-baryon pairs are small, is found to be attractive in all distances and
produces a quasi-bound state 1.54 MeV for $n\Omega ^{-}(uddsss)$ and 2.46
MeV for $p\Omega ^{-}(uudsss)$. In the latter case, the BE increase is due to the extra Coulomb attraction. The fitted lattice QCD
potential was constructed by Gaussian and Yukawa squared form for obtained observables such
as the scattering phase shifts, root-mean-square distance, and binding
energy.
The adjustment parameters are found from the
simulation 
and this potential 
has the form \cite{Iritani2019}:
\begin{equation}
V_{\Omega N}=b_{1}e^{-b_{2}r^{2}}+b_{3}\left( 1-e^{-b_{5}r^{2}}\right)
\left( \frac{e^{-m_{\pi }r}}{r}\right) ^{2},  \label{HALInteraction}
\end{equation}%
where $b_1$=-306.5 MeV, $b_2$=273.9 fm$^{-2}$, $b_3$=-266 MeV·fm, $b_4$=0.78 fm$^{-2}$. The
resultant scattering characteristics obtained with sets of parameters are
found to be consistent with each other within statistical errors. The Yukawa
squared form at long distances is motivated by the two-pion exchange between $N$ and $\Omega $.

In the single folding model the $\Omega$-$\alpha$ potential, $V_{\Omega\alpha}(r)$, can be obtained as \cite{Satchler1983}:
\begin{equation}
\label{folding}
V_{\Omega\alpha }(r)=\int \rho (\mathbf{x)}V_{\Omega N}(\mathbf{r-x)}d\mathbf{x},
\end{equation}
where  $\rho (\mathbf{x)}$ is the density of nucleons in $^{4}$He and $\left\vert \mathbf{r-x}\right\vert $ is the distance between the $\Omega$ baryon and nucleon. 
At large distances between $\Omega$ and $\alpha$-particle the clustering is described as $\Omega +(NNNN)$ system.
In the region near the $\alpha$-cluster and inside, the clustering must include different combinations $(\Omega NNN)+N$,
$(\Omega NN )+(NN)$, $(\Omega N)+(NNN)$. 
We assume that the $\Omega+(NNNN)$ clusterization dominates outside of the $\alpha$-particle. 

The effective $\Omega$-$\alpha$ interaction was obtained using a folding potential method \cite{Satchler1983,Miyamoto2018} and the central HAL QCD potential, and it is based on the $\Omega$-$N$ potential  \cite{Iritani2019} which is fitted to low energy $\Omega N$
scattering parameters.  Considering the central symmetry of the $V_ {\Omega N}(r)$ potential (\ref{HALInteraction}) and density $\rho(r)$, expression (\ref{folding}) reads
\begin{equation}
V_{\Omega\alpha}(r)=\int_{-1}^1du\int_0^{R_{max}}dx \rho(x)V_{\Omega-N}(\sqrt{x^2+r^2-2xru})x^2,
\end{equation}%
where $\rho(x)$ represents the density function of a nucleon in the $\alpha$-particle, normalized by the condition: 
$\int\rho(\mathbf{r)}d\mathbf{r} = 4$. Dover and Gal \cite{Gal83} proposed for the $\Xi$-$\alpha$ interaction the WS type potential. Following \cite{Gal83,GHM}, we fit the folding potential using a simple WS type
expression
\begin{equation}
V(r)=V_{0}\left[ 1+\exp \left( \frac{r-R}{c}\right) \right]^{-1},
\label{Omegaalfa}
\end{equation}%
assuming to be derived from the asymptotic region $r>R_{rms}$, where $R_{rms} = \left\langle r^{2}\right\rangle ^{1/2}$ is the root-mean-square ($rms$) radius of $\alpha$-particle. This region was chosen according to the core radius of the density function of a nucleon in the $\alpha$-particle due to the quick decreasing of the density function along the distance between $\Omega$ and $\alpha$-particle.

The $R_{rms}$ radius is an important and basic property for $^4$He, and for this nucleus both the matter and charge $rms$ radii were measured. A matter radius is related to both the proton and the neutron distributions inside a nucleus, whereas a nuclear charge radius is primarily connected to the proton distribution. The average $rms$ charge radius of $^4$He from electron elastic scattering experiments is 1.681(4) fm \cite{Sick2008}. 
The result of precise measurements of  $^4$He $rms$ charge radius with the technique of muon-atom spectroscopy gives 1.67824(13) fm \cite{Krauth}. While recently, by examining the near-threshold $\phi$-meson photoproduction data of the LEPS Collaboration \cite{HiraiwaLEPS}, the $rms$ matter radius of $^4$He is measured to be 1.70 ± 0.14 fm \cite{Wang2024}.  From these analyses, the $rms$ charge radius of $^{4}$He is smaller than the $rms$ matter radius. However, the values of the $rms$ charge and matter radii are within the statistical errors. 
Whereas, in our calculation of the folding potential we used the density that reproduces the $rms$ radii 1.70 ± 0.14 fm \cite{Wang2024}. It yields the minimum and maximum values of  1.56 fm and 1.84 fm. The latter allows us to study the influence of the $rms$ on the $\Omega$-$\alpha$ potential.
As follows from Ref. \cite{Wang2024} the simple Gaussian matter distribution model 
\begin{equation}
\rho_1 (r)=\left( {C^{2}}/{\pi }\right) ^{3/2}e^{-C^{2}r^{2}}
\label{Rho1}
\end{equation}%
gives $R_{rms}=\sqrt{3/2C^{2}}$ and describes the experimental data with parameters from \cite{Wang2024}. The matter density distribution in $^4$He that leads to the experimental $rms$ radius of  1.70 fm is shown  Fig. \ref{fig:06}(a). Another model of the matter density distribution proposed in Ref. \cite{Thomas97} was chosen to reproduce the $rms$ matter radius of $^4$He and the measured central depression in the charge density has the form 
\begin{equation}
\rho_2 (r)=A(1+\alpha r^{2})\exp (-\beta r^{2}).
\label{Rho2}
\end{equation}%
In Eq. (\ref{Rho2}) parameters $A$=0.084 MeV, $\alpha =$1.34215 fm$^{-2}$ and $\beta =$0.904919 fm$^{-2}$
were chosen for the $rms$ matter radius of $^{4}$He, 1.56 fm. The matter density distribution
in $^4$He that leads to the experimental $rms$ within the experimental statistical uncertainty is shown in the insert in Fig. 1($a$).

\section{Results of calculations and discussion}

\label{Calculations}

\begin{figure}[ht]
\begin{center}
\includegraphics[width=20.5pc]{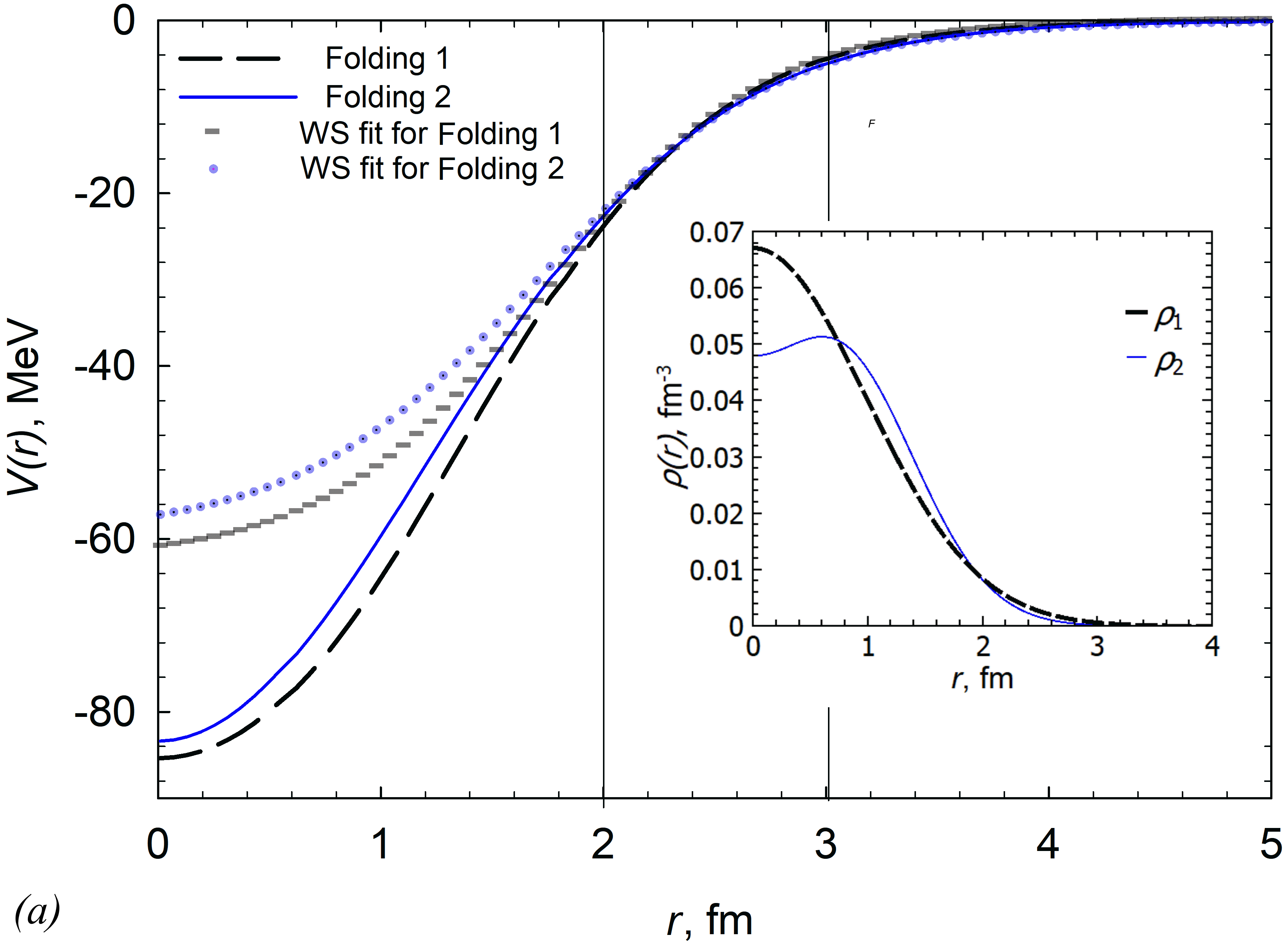}
\includegraphics[width=19.5 pc]{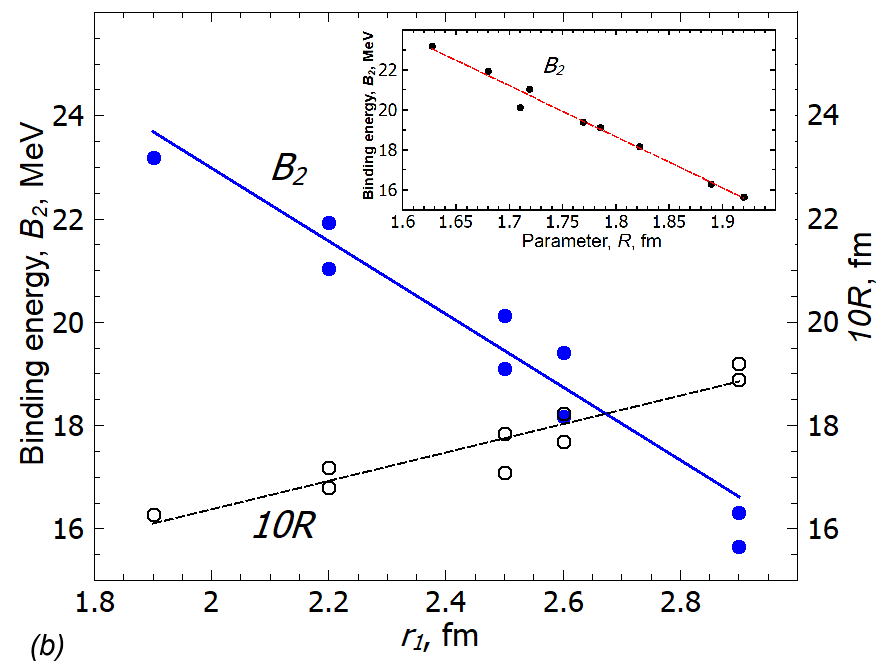}
\end{center}
\caption{($a$) The folding $\Omega$-$\alpha$ potentials as a function of the distances between particles $r$: Folding 1 (black dashed curve) and Folding 2 (solid blue curve). A dashed and dotted curves show the WS fits for these potentials. The $rms$ radius of the $\alpha$-particle is chosen to be 1.7~fm for the Folding 1 and 1.56 fm for the Folding 2 potentials. Insert: The matter density distribution $\rho(r)$ in $^4$He corresponding to the model functions (\ref{Rho1}) and (\ref{Rho2}). The dashed curve  
corresponds to $\rho_1(r)$ and the experimental $R_{rms} = 1.70$~fm. 
The solid curve corresponds to $\rho_2$  and $R_{rms}=1.56$~fm.
($b$) The dependence of the $\Omega$+$\alpha$ system binding energy and the WS parameter $R$ on the choice of $r_1$ point. Inset: The relation between $B_2$ binding energy (solid circles) of the $\Omega$+$\alpha$ system and the WS parameter $R$ for the Folding 1 potential.
The solid and dashed lines are a linear fit for the calculated data.
}
\label{fig:06}
\end{figure}

We utilized the single folding model with the different mass distribution functions and the HAL QCD $V_{\Omega N}$ potential to construct a $V_{\Omega \alpha}$ potential. The folding $\Omega$-$\alpha$ potentials presented in Fig. \ref{fig:06}($a$) were calculated using different nucleon density functions, represented by Eqs. (\ref{Rho1}) and (\ref{Rho2}). The root-mean-square radius was set to 1.7 fm for the Gaussian
mass distribution model (\ref{Rho1}) and $R_{rms} = 1.56$ fm for the density function (\ref{Rho2}). In the insert, the matter density distribution $\rho(r)$ in $^4$He is shown, corresponding to the different models. The solid curve, $\rho_1(r)$, corresponds the experimental $rms$ value of 1.70 fm,  
while the dashed curve, $\rho_2(r)$, represents the $rms$ value of 1.56 fm. The corresponding $rms$ is obtained using Eqs. (\ref{Rho1}) and (\ref{Rho2}). The resulting potentials are referred to as the Folding 1 and Folding 2 potentials, respectively.
We also depict the Woods-Saxon fit 
\begin{equation}
V_{\Omega\alpha}(r_i)=V_{0}\left[ 1+\exp \left( \frac{r_i-R}{c}\right) \right]^{-1}, \quad i=1,2,3.
\label{set}
\end{equation}
for both folding potentials shown in Fig. \ref{fig:06}($a$). The WS potential parameters $V_0$ (the strength of the interaction), $c$ (the surface diffuseness), and $R$ were determined by numerically solving a set of nonlinear equations.
We developed a Python code using the pre-built function fsolve, which iteratively refines guesses for the solution by leveraging the Jacobian matrix of the system to guide the search toward the roots. In Ref. \cite{Etminan2020}, the WS fit for the $\Omega$-$\alpha$ potential was constructed for the interval $r > 2$ fm, using the initial parameters from the known Dover and Gal (DG) \cite{Gal83} $\Xi$-$\alpha$ potential. For our fit, we used the range $r$ from 2 to 3 fm with similar initial parameters: $V_0 = -27$ MeV, $R = 1.6$ fm, and $c = 0.5$ fm. The mesh of coordinates $r_i$ was selected within the region $1.9 < r_i < 3.2$ fm.
\begin{figure}[ht]
\begin{center}
\includegraphics[width=18pc]{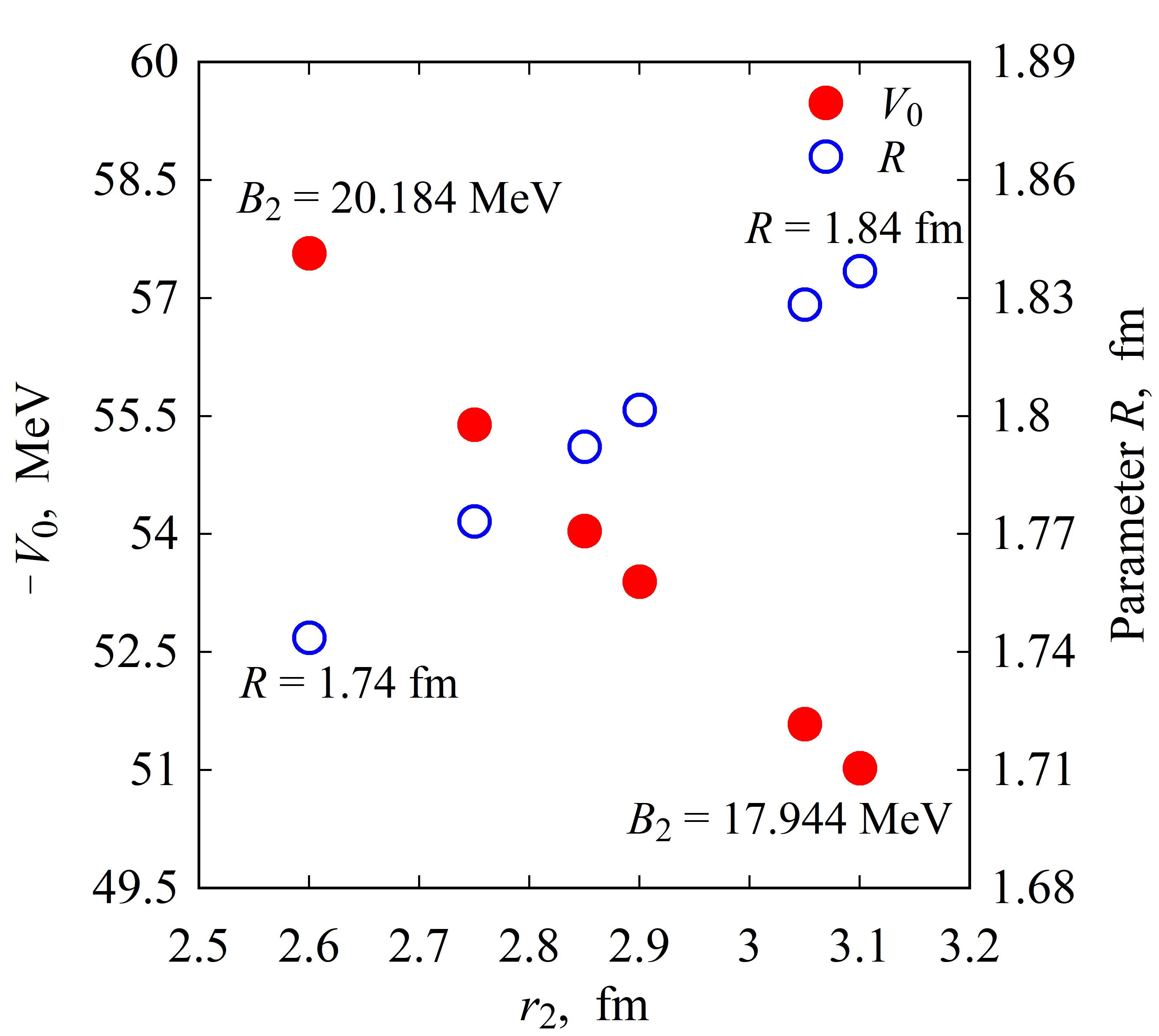}
\end{center}
\caption{ The parameters $V_0$ (solid circles) and $R$ (open circles) of the WS fit of the $\Omega$-$\alpha$ Folding 1  potential for diffrenet values of $r_2$, when $r_1$=2.5 fm and $r_3$=3.2 fm. 
We show the values of the $\Omega$+$\alpha$ system binding energy $ B_2$ and the WS parameter $R$ for two cases: $r_2$=2.6 fm and $r_2$=3.1 fm.}
\label{fig:07}
\end{figure}

The numerical results related to the folding procedure are presented in Table \ref{t1} for several sets of coordinates $r_i$ ($i=1,2,3$). For each coordinate set, we obtained a WS fit, defined by three Woods-Saxon parameters $V_0$, $R$, and $c$. Additionally, in Table \ref{t1} we provide the binding energy $B_2$ of the $\Omega$+$\alpha$ system ($^{4}_{\Omega}$He hypernucleus), calculated using each WS fit. The analysis of the results in Table \ref{t1} shows significant variations in both the WS parameters and the binding energy, depending on the chosen coordinate set.
\begin{table}[!ht]
\caption{
The parameters  $V_0$, $R$ and $c$ of WS fits for the folding $V_{\Omega\alpha}(r)$ potential calculated for two types of the density function $\rho(r)$ (Folding 1 and Folding 2).  
$B_2$ is the two-body bounding energy of the $\Omega$+$\alpha$ system. The WS parameters chosen for the  Folding 1 and Folding 2 potentials are marked by stars.
}
\label{t1}
\begin{tabular}{cccccccc} \hline\noalign{\smallskip}
$i$&$V_{\Omega-\alpha}(r_i)$, MeV &$r_i$, fm& $V_0$, MeV& $R$ (fm) & $c$, fm& $B_2$, MeV\\
\hline\noalign{\smallskip}
1&-25.49546 &1.9  &  -88.92323& 1.36075 & 0.59166&28.096\\
2&-22.53559&2.0  &  &  & \\
3&-19.81202&2.1  &  &  & \\
1&-25.49546 &1.9  &  -65.783803& 1.65274 & 0.540371&23.184\\
2&-5.95020&2.9  &  &  & \\
3&-4.2278&3.1  &  &  & \\
1&-17.3279 &2.2   & -62.70735& 1.68007 & 0.54005&21.940\\
2&-9.65817&2.6  &  &  & \\
3&-3.54861&3.2  &  &  & \\
1&-17.3279 &2.2   & 59.79104& 1.719691 &0.53586&21.046\\
2&-5.95020&2.9  &  &  & \\
3&-4.2278&3.1   &  &  & \\
1&-11.2537 &2.5   & -59.53264& 1.71473 & 0.53929& 20.137$^{*}$\\
2&-8.25138&2.7  &  &  & \\
3&-5.95020&2.9  &  &  & \\
1&-11.2537 &2.5   & -54.4667& 1.785629& 0.53103& 19.113\\
2&-5.95020&2.9  &  &  & \\
3&-4.2278&3.1  &  &  & \\
1&-9.65817 &2.6   & -51.82068& 1.82253 & 0.52754&18.172\\
2&-5.9502&2.9  &  &  & \\
3&-3.546071&3.2  &  &  & \\
1&-9.65817 &2.6   & -55.4762& 1.769492 & 0.533445&19.412\\
2&-7.02062&2.8  &  &  & \\
3&-5.95020&3.0 &  &  & \\
1&-5.95020&2.9   & -45.13086& 1.920287 & 0.519807& 15.654\\
2&-4.2278&3.1  &  & \\
3&-3.5461&3.2  &  &  & \\
1&-5.95020&2.9   & -47.12945& 1.8895 & 0.52258&16.316\\
2&-5.02438&3.0  &  & \\
3&-4.15472&3.1  &  &  & \\ \hline\noalign{\smallskip}
1&-11.12 &2.49   & -62.714& 1.744 & 0.4860&23.751$^{**}$\\
2&-7.29&2.73  &  &  & \\
3&-4.66&2.97  &  &  & \\
\noalign{\smallskip}\hline \noalign{\smallskip}
\end{tabular} \\
\hspace{-4.7cm}
\begin{minipage}{10cm}
 Folding 1$^{*}$,  \quad Folding 2$^{**}$ 
 \end{minipage}
\end{table}
The coordinate sets differ based on the selection of different starting points, $r_1$, and ending points, $r_3$, as well as the placement of the intermediate point, $r_2$. Here, $r_1>1.9$ fm, and the sizes of the intervals $r_3 - r_1$ also vary. However, general conditions can be formulated as follows: the interval $r_3 - r_1$ should be approximately 1 fm, and $r_2$ should be positioned midway between $r_1$ and $r_3$.
For example, a comparison of the first and second sets, which differ by the values of $r_3 - r_1$, shows that this leads to substantial differences in both the binding energy and the WS parameters. We can consider these differences as an uncertainty in the folding procedure.
The criterion for selecting the coordinate interval is related to the WS range parameter $R$, which should be approximately 1.7 fm. We assume that $R$ is a phenomenological constant defined as 
$R = R_0 A^{1/3}$, with $R_0=1.2\pm 0.1$ fm.
We consider the following conditions for choosing the coordinates $r_1$ and $r_3$: 1.9 $< r <$ 3.2 fm. This defines the asymptotic region for the $\Omega$-$\alpha$ potential, where the dominant interaction is between $\Omega$ and $\alpha$-particle, and other possible channels, such as $\Omega NN$-$2N$, are neglected. 
Additionally, the lower bound $r_1$ must be larger than the $rms$ radius of the $\alpha$-particle, meaning $r_1 > 1.7$ fm, and more precisely, $r_1 > 1.85$ fm, to account for the uncertainty in experimental data for the radius.

According to our criteria, the binding energy of the bound $\Omega \alpha$ pair was calculated to be approximately 20 MeV, as shown in Table \ref{t1}. This result is close to the one obtained in Ref. \cite{Etminan2020}, but the discrepancy suggests that the folding procedure is not entirely reliable and is sensitive to the input values. However, it is important to note that fitting across the entire range of distances $r_1 - r_3$ exhibits stability. Deviations in the parameters used for adjusting the folding potential with varying levels of accuracy do not significantly alter the binding energy of the $\Omega$+$\alpha$ system.

The deviations are graphically presented in Fig. \ref{fig:06}(b). Notable,  the calculated data is generally well-fitted by a linear function. The small variations, denoted by \( \delta \) in the potential $ V_{\Omega\alpha}(\mathbf{r})$ in the Schr\"odinger equation
\begin{equation}
\left( -\frac{\hbar^2}{2m} \nabla^2 \Psi(\mathbf{r}) + V_{\Omega\alpha}(\mathbf{r}) + \delta \right) \Psi(\mathbf{r}) = (E + \epsilon) \Psi(\mathbf{r}),
\end{equation}
where 
$m$ is the reduced mass of the $\Omega$+$\alpha$ pair,
$\nabla^2$ is the Laplace operator, 
$V_{\Omega\alpha}(\mathbf{r})$ is the potential, 
$E$ is the energy of the ground state, and 
 $\Psi(\mathbf{r})$ is the wave function of the state that 
leads to a small correction \( \epsilon \) in the energy. 
In our case, the variation \( \epsilon \) arises from changes in the set of parameters of the $\Omega$-$\alpha$ Woods-Saxon potential, as shown in Table \ref{t1}. The variations in the binding energy are approximately 1-2 MeV, as seen in Fig. \ref{fig:06}($b$). Consequently, the relative changes in the binding energy are less than 10\%. According to the first order of the perturbation theory, this dependence can be represented as linear, and our calculation results confirm this relation.

A linear dependence is also observed in the relationship between the Woods-Saxon parameters obtained by varying the midpoint $r_2$ of the mesh $r_i$ ($i=1,2,3$). In Fig. \ref{fig:07}, the parameters $V_0$ and $R$ of the WS fit for the Folding 1 $\Omega$-$\alpha$ potential are shown for different values of $r_2$. The interval $r_3 - r_1$ was chosen with $r_1 = 2.5$ fm and $r_3 = 3.2$ fm. Within this range, the binding energy of the $\Omega$+$\alpha$ system varies from 18 MeV to 20 MeV.

The $ \Omega$+$\alpha$ system bound
state with a binding energy of approximately $22$ MeV was evaluated in Ref. \cite{Etminan2020}  using the parameterization with
 a strength $V_{0}=-61$ MeV, surface diffuseness $c=0.47$, and  radius $R=1.74$ fm. 
It is important to note that, due to the lack of sufficient experimental data for the $^5_{\Omega}$He bound state, the depth of the potential remains a hypothetical value.
\begin{table}[!ht]
\caption{
The low-energy characteristics of the $Y$-$\alpha$ potentials. The $\Omega$-$\alpha$ potential was obtained using the folding procedure with different root mean square radii for the matter density function of the $\alpha$-particle, as described by Eq. (\ref{Rho1}) (Folding 1) and Eq. (\ref{Rho2}) (Folding 2). The calculated and experimental binding energy of a $Y\alpha$ pair marked as $B_2$ and $B_2^{exp}$, respectively.
}
\label{t00}
\begin{tabular}{ccccccl} \hline\noalign{\smallskip}
Potential& $rms$, fm &Nucleus& $a_{Y \alpha}$, fm & $r_{Y \alpha} $, fm& $B_2$, MeV &$B_2^{exp}$, MeV \\
\hline\noalign{\smallskip}
$\Omega$-$\alpha $ Folding \cite{Etminan2020}& 1.47  &$^5_\Omega$He & 1.0  & 1.1& 22.01 &\\
           $\Omega$-$\alpha $                             Folding 2   & 1.56 &      $^5_\Omega$He    &0.46   &  14.6& 23.751& \\
           $\Omega$-$\alpha $                             Folding 1    & 1.71  &      $^5_\Omega$He  &  0.36   & 35.0 & 20.137 &\\
\hline \noalign{\smallskip} 
 $\Lambda$-$\alpha$ \cite{Isle} (Isle)&--& $^5_\Lambda$He&4.2 &1.9& 3.10 \cite{FSV09}&3.12$\pm$0.02 \cite{GHM}\\
$\Xi $-$\alpha$ \cite{Gal83}&--&$^5_\Xi$He  & 4.6 & 2.2 & 2.09 \cite{FSV08}& \\
 \hline \noalign{\smallskip} 
\end{tabular}
\end{table}

The results for the ground state energy of the ${^{5}_{\Omega}} \text{He}$ hypernucleus, obtained using the Woods-Saxon fits of the folding $\Omega$-$\alpha$ potential, are presented in Table \ref{t00}. The low-energy $\Omega$+$\alpha$ system characteristics exhibit a significant dependence on the $rms$ radius utilized in the corresponding folding procedure. A more appropriate value for the $rms$ radius, $1.7 \, \text{fm}$, yields a larger effective scattering radius $r_{\Omega\alpha}$. This suggests that the energy dependence of the scattering phase shift $\delta(k)$ is notable:
 $
 k\text{cot} \delta(k)=- 1/a + \frac{k^2}{2}r_{eff} + \dots,
$
where $k$ is the wave number. This dependence distinguishes the folding potentials described in \cite{Etminan2020} from the Folding 1 model. Additionally, a comparison with previous calculations for the $\Lambda$+$\alpha$ ($^4_\Lambda\text{He}$) and $\Xi$+$\alpha$ (${^{4}_{\Xi}}\text{He}$) systems reveals significant discrepancies in the effective radius and binding energy. 
It is important to note that for the $^5_\Lambda\text{He}$ nucleus there is experimental data presented in Table. \ref{t00}. The experimental binding energy $B_2^{exp}$ for $^5_\Lambda\text{He}$ is approximately $3 \, \text{MeV}$, which contrasts sharply with the value of $20 \, \text{MeV}$ for the $^5_\Omega\text{He}$ hypernucleus.

\begin{figure}[ht]
\begin{center}
\includegraphics[width=20pc]{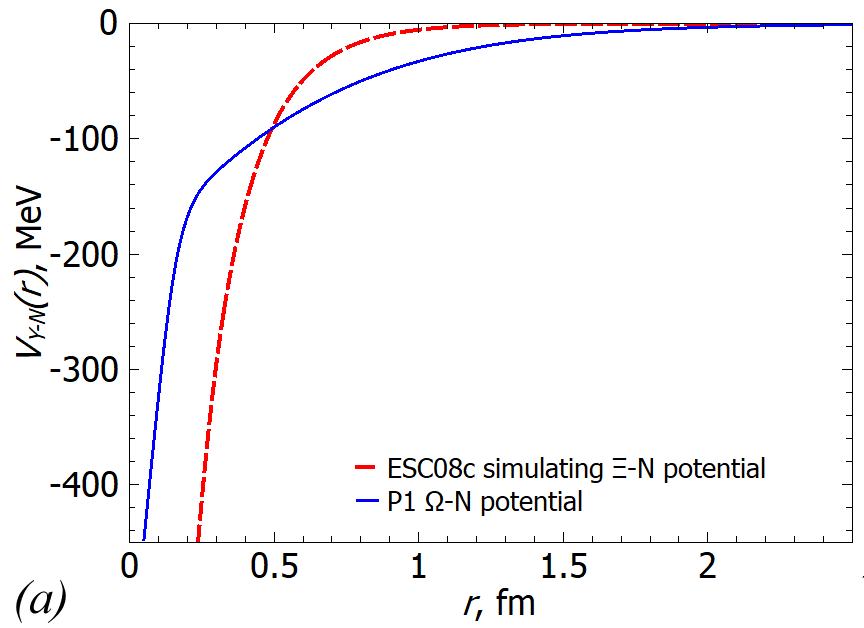}
\includegraphics[width=20pc]{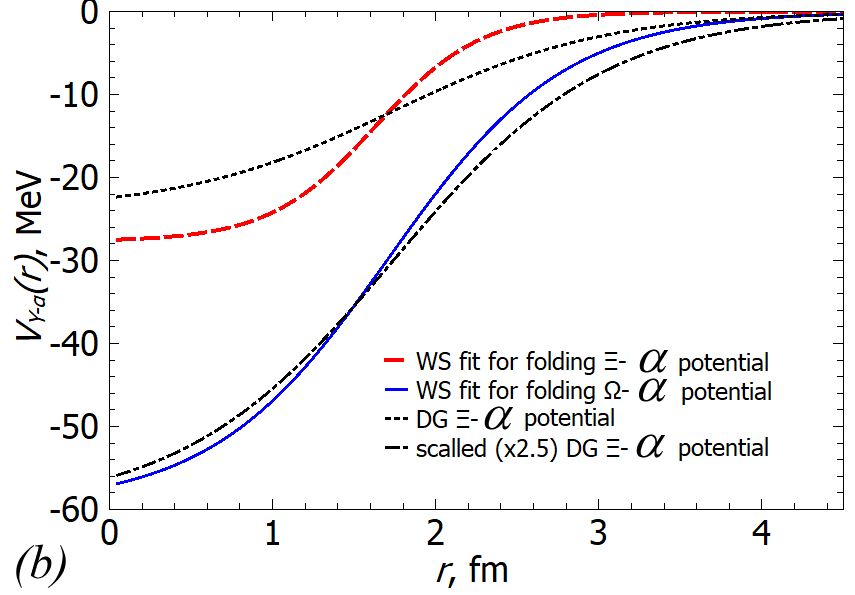}
\end{center}
\caption{ ($a$) The simulating  ESC08c $\Xi$-$N$ (dashed curve) and $\Omega$-$N$ with parameters $P1$ \cite{Iritani2019} (solid curve) potentials. ($b$) The WS fit for folding $\Xi$-$\alpha$ (dashed curve) and $\Omega$-$\alpha$ (solid curve) potentials. The dotted and dash-dotted curves show DG $\Xi$-$\alpha$ \cite{Gal83} and scaled (x2.5) DG $\Xi$-$\alpha$ potentials.
}
\label{fig:08}
\end{figure}

To evaluate the described folding procedure, we use the $\Xi$-$N$ potential which simulates
the ESC08c $Y$-$N$ Nijmegen model \cite{ESC08c} by two Yukawa-type functions with attractive and repulsive
terms \cite{GVV}
$$
V_{\Xi N}(r)= (-568exp(-4.56r)+425exp(-6.73r))/r.
$$ 
Following Ref. \cite{Etminan2020}, we construct a $\Xi$-$\alpha$ folding potential using the same method applied to the $\Omega$-$\alpha$ potential.  The $Y$-$N$ potentials for $Y=\Xi$, $\Omega$ are illustrated in Fig. \ref{fig:08}(a) as functions of the relative coordinate $r$. A comparison reveals that the $\Omega$-$N$ potential is a larger magnitude and has a longer tail at distances $r>$0.2 fm. In other words, the $\Xi$-$ N$ potential is a shorter range potential. This situation is sensitive to variations in the coefficient $b_5$ in Eq. (\ref{HALInteraction}).
For values of $b_{5}$ less than 0.78 fm$^{-2}$, the tail of the potential is both reduced in magnitude and range.

The numerical results of the folding procedure for the $\Xi$-$\alpha$  potential are presented in Table \ref{t3}. The criteria for selecting the parameters of the Woods-Saxon fit for the $\Xi$-$\alpha$  potential is again based on the condition $R \approx$ 1.7 fm. The resulting folding potential, with WS parameters similar to those in Ref. \cite{Etminan2020}, is labeled as Folding 3. In Ref. \cite{Etminan2020}, these parameters are calculated with the Gaussian matter distribution nuclear density of the alpha particle, yielding the $rms$  radius of 1.47~fm .
In Table \ref{t3}, Folding 4 corresponds to the $rms$  radius of 1.7~fm, while Folding 5 corresponds to the $rms$  radius of 1.84~fm. It is important to note that the WS parameters are similar across these three cases. However, there is a tendency for the diffuse parameter $c$ to increase with the $rms$ radius. The parameter $c$ defines the decrease of the folding potential at large distances.
The results for the folding potentials of the $Y$-$\alpha$ interactions are illustrated in Fig. \ref{fig:08}($b$).
\begin{table}[!ht]
\caption{
The parameters  $V_0$, $R$ and $c$ of WS fits for folding $V_{\Xi\alpha}(r)$ potential calculated for the Gaussian matter distribution functions $\rho(r)$ (\ref{Rho1}) with parameters giving the $rms$ radius 1.47 fm (Folding 3), 1.7 fm (Folding 4), and 1.84 fm (Folding 5).  
$B_2$ is the $\Xi$+$\alpha$ system binding energy.
}
\label{t3}
\begin{tabular}{cccccccc} \hline\noalign{\smallskip}
$i$&$V_{\Xi-\alpha}(r_i)$, MeV &$r_i$, fm& $V_0$, MeV& $R$ (fm) & $c$, fm& $B_2$, MeV\\
\hline\noalign{\smallskip}
1&-6.7022 &2.0  &  -18.951& 1.8192& 0.29979&0.57273\\
2&-1.7733&2.5  &  &  & \\
3&-0.36204&3.0  &  &  & \\
--&--&-- &  -24.4& 1.72 & 0.31&1.54 \cite{Etminan2020}\\
1&-6.7022 &2.0  &  -27.706& 1.6292 & 0.32458&1.8910$^{***}$\\
2&-3.5659&2.25  &  &  & \\
3& -1.7733&2.5  &  &  & \\
1&-6.7022&2.0   & -33.263& 1.5364 & 0.33663&2.8636\\
2&-4.6277&2.15  &  &  & \\
3&-3.5659&2.25  &  &  & \\  \hline\noalign{\smallskip}
--&--&-- &  -24& 1.74 & 0.65&2.09 \cite{FSV17}\\
1&-7.8511 &2.0  &  -25.1707& 1.6973 & 0.38251&1.6755$^{IV}$\\
2&-4.31518&2.3  &  &  & \\
3& -1.3345&2.8 &  &  & \\ \hline\noalign{\smallskip}
1&-6.8792 &2.1  &  -23.018& 1.7407 & 0.42128&1.3496$^{V}$\\
2&-4.3151&2.3  &  &  & \\
3& -1.3808&2.9 &  &  & \\
\noalign{\smallskip}\hline \noalign{\smallskip}
\end{tabular} \\
\hspace{-2.8cm}
\begin{minipage}{10cm}
  Folding 3$^{***}$, \quad Folding 4$^{IV}$, \quad Folding 5$^{V}$
 \end{minipage}
\end{table}
 Here, we demonstrate a difference between the phenomenological $\Xi$-$\alpha$ DG \cite{Gal83}  and folding $\Xi$-$\alpha$ potentials. One can see from the table that the folding procedure, which involves simulating the ESC08c $\Xi$-$N$ \cite{ESC08c} potential, cannot achieve the parameters of the phenomenological $\Xi$-$ \alpha$  potential. The folding potential is noticeably deeper.
This is not surprising, as the folding procedure must generally be accompanied by a scaling \cite{double} to reproduce experimental data. In this case, however, the folding $\Xi$-$\alpha$ potential cannot be simply scaled to reproduce the DG potential parameters, which are considered to be "experimentally motivated" \cite{Etminan2020}. In particular, the parameter $c$ is essentially small for the Woods-Saxon  fit of the $\Xi$-$\alpha$ folding potential.
Note here the paradoxical scaling of the $\Xi$-$\alpha$ DG potential, which closely approaches the folding $\Omega$-$\alpha$ potential, as shown in Fig. \ref{fig:08}(b). The small parameter $c$ corresponds to a rapid decrease in the $\Xi$-$\alpha$ folding potential in the asymptotic region, which is restricted by a region near the alpha particle radius. Thus, the choice of a "good" asymptotic region becomes difficult due to unstable results for the WS parameters (see the Folding 3 in Table \ref{t3}).
We can conclude that the $\Xi$-$N$ potential used in this study generates a small $c$ parameter and an asymptotic region that is too short for a reliable folding procedure involving the Woods-Saxon function.

\section{Concluding Remarks}
\label{Summary}
We obtain $\Omega$-$\alpha$ potential following the folding procedure proposed in Refs. \cite{Etminan2020,FilKez2024}. We repeated this procedure to confirm its suitability for the describtion of the $\Omega$-$\alpha$ interaction. The resulting binding energy, $B_2$, of the $\Omega$+$ \alpha$ system is approximately 20 MeV, consistent with the results from Ref. \cite{Etminan2020}. This significant binding energy of the hypothetical nucleus $^{5}_{\Omega}$He can primarily be attributed to the strong attractive $\Omega$-$N$ potential, which yields a binding energy 1.54 MeV of the $\Omega N$ dibaryon.

The folding $\Omega$-$\alpha$ potential, $V_{\Omega\alpha}(r)$, was fitted using the Woods-Saxon formula, which involves three free parameters: $V_0$, $R$, and $c$. The Woods-Saxon model closely aligns with the strong attractive $\Omega$-$ N$ potential. However, the binding energy of the hypothetical $^{5}_{\Omega}$He hypernucleus is about ten times greater than the binding energies of similar hypernuclear systems, such as $^{5}_{\Lambda}$He and $^{5}_{\Xi}$He. The purely attractive nature of the obtained folding potential may contribute to the large $\Omega$+$\alpha$ binding energy. In contrast, the $\Lambda$-$\alpha$ and $\Xi$-$\alpha$ potentials proposed in other studies include a repulsive core, resulting in weaker binding of the $^{5}_{\Lambda}$He and $^{5}_{\Xi}$He.

This large $B_2$ may be related to the method used to construct the $\Omega$-$\alpha$ potential. We used the following conditions as inputs for the construction: i. we considered the WS range parameter $R$ to be approximately 1.7 fm as a phenomenological constant, where $R$ is defined as $R = R_0A^{1/3}$, with $R_0$=1.2$\pm$0.1 fm; ii. the folding procedure applies to a limited range of the variable $r$, specifically $1.9 < r < 3.2$ fm, considered the asymptotic region for the $\Omega$-$\alpha$ potential. In this region, the dominant channel is the $\Omega$+$\alpha$ channel, while possible channels such as $(\Omega NN)$-$2N$ and others are ignored. This introduces uncertainty in the folding potential calculated in our study and Ref.~\cite{Etminan2020}. Note that by choosing the lower range limit of 2~fm, we ensured that this range is larger than the $rms$ radius of the alpha particle.
In this context, variations in the nucleon density of the $\alpha$-particle within the folding integration do not significantly affect the calculated $\Omega$+$\alpha$ system binding energy. The change of the BE is in the range of few MeV. 

We identified another source of uncertainty related to the choice of the mesh points $r_i$ ($i = 1, 2, 3$) used in the WS fitting. Variations in the position of $r_2$ within the range $2 < r < 3$ can visibly change the WS parameters, affecting the $\Omega$+$\alpha$ binding energy by 1–2 MeV. In particular, we observed a linear dependence of $B_2$ on $R$.

We performed folding calculations for the $\Xi$+$\alpha$ system using the $\Xi$-$ N$ potential, which simulates the ESC08c $Y$-$N$ Nijmegen model through two Yukawa-type functions with attractive and repulsive terms \cite{GVV}. 
This potential generates unstable calculations for the folding $\Xi$-$\alpha$ potential which is caused by a short asymptotic region for the folding procedure and results in a small WS parameter $c$.  Consequently, the procedure does not lead to a reliable evaluation of the $\Xi$-$\alpha$ interaction.

Our general conclusion is that the folding procedure for the $\Omega$-$\alpha$ potential is accompanied by some uncertainties, and the large binding energy of the $\Omega \alpha$ pair is not yet a reliable value. The main uncertainty is related to the assumption that the $\Omega$+$\alpha$ channel dominates in the asymptotic region. Further investigation into $\Omega$-$N$ interactions is necessary to clarify the behavior of the $\Omega$-$\alpha$ potential at short distances. 

It is also worth noting that new experimental facilities dedicated to the study of $\Omega$ baryons are expected to be developed in the near future \cite{S21}.
\section*{Acknowledgments}
We would like to express our gratitude to P. Bautista for help with some  
calculations.
R.Ya.K. is grateful for the support by the City University of New York PSC CUNY Research Award \# 66109-00 54, I.F. and B.V are supported by the DHS Science and Technology Directorate Office of University Programs Summer Research Team Program for Minority Serving Institutions through an interagency agreement between the U.S. DOE and DHS. ORISE is managed by ORAU under DOE contract number DE-SC0014664. All opinions expressed in this paper are the author's and do not necessarily reflect the policies and views of DHS, DOE or ORAU/ORISE. This work is partly supported by US National Science Foundation HRD-1345219 award and the Department of Energy/National Nuclear Security Administration Award \# NA0004112.

\end{document}